\documentclass[aps,prl,twocolumn,showpacs,superscriptaddress]{revtex4-2}
\usepackage{graphicx}  % needed for figures
\usepackage{dcolumn}   % needed for some tables
\usepackage{bm}        % for math
\usepackage{amssymb}   % for math
\usepackage{xcolor}
\usepackage{hyperref}
\usepackage{amsmath}

\begin{document}

%\title{Polar vs chiral magneto-structural coupling in cubic quadruple perovskites}

\title{Polarity vs Chirality: Functionality from competing magneto-structural instabilities}

\author{M. Tardieux}
\author{E. Stylianidis}
\author{D. Behr}
\affiliation{Department of Physics and Astronomy, University College London, Gower Street, London WC1E 6BT, United Kingdom}

\author{R. Liu}
\author{K. Yamaura}
\affiliation{Research Center for Materials Nanoarchitectonics (MANA), National Institute for Materials Science (NIMS), 1-1 Namiki, Tsukuba, Ibaraki 305-0044, Japan}
\affiliation{Graduate School of Chemical Sciences and Engineering, Hokkaido University, North 10 West 8, Kitaku, Sapporo, Hokkaido 060-0810, Japan}

\author{D. D. Khalyavin}%
\affiliation{ISIS facility, Rutherford Appleton Laboratory-STFC, Chilton, Didcot OX11 0QX, United Kingdom}

\author{D. Bowler}
\affiliation{Department of Physics and Astronomy, University College London, Gower Street, London WC1E 6BT, United Kingdom}
\affiliation{London Centre for Nanotechnology, 17-19 Gordon St, London WC1H 0AH, United Kingdon}
\affiliation{Research Center for Materials Nanoarchitectonics (MANA), National Institute for Materials Science (NIMS), 1-1 Namiki, Tsukuba, Ibaraki 305-0044, Japan}

\author{A. A. Belik}
\affiliation{Research Center for Materials Nanoarchitectonics (MANA), National Institute for Materials Science (NIMS), 1-1 Namiki, Tsukuba, Ibaraki 305-0044, Japan}

\author{R. D. Johnson}
\email{roger.johnson@ucl.ac.uk}
\affiliation{Department of Physics and Astronomy, University College London, Gower Street, London WC1E 6BT, United Kingdom}

\date{\today}

\begin{abstract}
We report a phenomenological magneto-structural model based on competing free-energy terms that couple either polar or chiral distortions in cubic quadruple perovskites, depending on the global direction of magnetic moments. The model naturally explains why some compounds in this material system host magnetically-induced ferroelectricity at low temperature, while others such as CaMn$_3$(Cr$_3$Mn)O$_{12}$, which we characterise experimentally, do not. Importantly, our results suggest a new approach towards developing an applied multiferroic functionality, and can be generalised to other multi-sublattice systems where the magnetic interaction between sublattices is prohibited by spatial inversion.

\end{abstract}

\maketitle

Coupling between structural distortions lies at the heart of many applicable material functionalities such as piezoelectricity and piezomagnetism, linear and non-linear magneto-electric effects, hybrid ferroelectricity, and optical effects associated with a lack of spatial inversion. In all such cases, physical properties can be anticipated by phenomenological symmetry analysis, making this technique a powerful tool for advancing our understanding and discovery of material functionality. One of the most pronounced examples of coupling phenomena determined by symmetry is the spin driven electric polarization in multiferroic materials. These systems are unusual as they combine at least two ferroic orders in the same phase, and they have garnered significant interest in physics and materials science due to their complex phenomenology and potential for application in new solid-state technologies \cite{Spaldin2010,Dong2015}. Numerous multiferroics have now been identified, and in almost all cases the microscopic interactions responsible for magnetoelectric coupling can be linked to one of three mechanisms; the Dzyaloshinskii-Moriya interaction \cite{Dzyaloshinsky1958,Kenzelmann2005,Mostovoy_2006,Kimura2007}, exchange-striction \cite{Choi_2008,Radaelli_2009,Mochizuki_2010,Behr_2023}, or $d$-$p$ hybridization \cite{Arima_2007}. Furthermore, several prototype devices have been produced and tested. However, the success of these devices appears to be limited by a fundamental constraint imposed by symmetry; an electric polarisation and ferromagnetic moment alone cannot be linearly coupled in a single phase. New approaches towards magneto-structural functionality, motivated by symmetry analysis and material characterisation, may therefore prove essential in realising future multiferroic devices.

In 2015, X. Wang \textit{et al.} \cite{Wang_2016} reported that the cubic quadruple perovskite LaMn$_3$Cr$_4$O$_{12}$ supported a magnetically driven electric polarisation that could not be explained by the above, well-established mechanisms. The quadruple perovskite AA'$_3$B$_4$O$_{12}$ has a cubic ($Im\bar{3}$) aristotype crystal structure \cite{Vasilev_2007}, which can be obtained by an ordered occupation of three-quarters of the simple perovskite (ABO$_3$) A-sites by a transition-metal ion, labelled A'. This cation order is stabilised through strong octahedral tilts ($a^+a^+a^+$ in Glazer notation \cite{Glazer1975}) that form square-planar A'O$_4$ units while maintaining the typical regular octahedral coordination of the B-sites. In many quadruple perovskites, cubic symmetry is broken at room temperature due to effects of charge order \cite{Bochu_1980}, the cooperative Jahn-Teller effect (orbital order) \cite{Prodi_2009}, or ferroelectric instabilities \cite{Mezzadri2009}. However, LaMn$_3$Cr$_4$O$_{12}$, with A = La$^{3+}$, A' = Mn$^{3+}$, and B = Cr$^{3+}$, remains cubic down to the lowest measured temperatures \cite{Wang_2016}. Two magnetic phase transitions were observed, the first at $T_\mathrm{N} = 150~\mathrm{K}$ is associated with B-site Cr$^{3+}$ G-type antiferromagnetic order, and the second at $T_2 = 50~\mathrm{K}$ marks the onset of A'-site Mn$^{3+}$ G-type antiferromagnetic order \cite{Lv_2011,Wang_2016}. In both cases the absolute direction of magnetic moments could not be determined owing to limitations inherent to neutron powder diffraction, but all ordered moments were found to be collinear in both phases. The unconventional magnetically driven electric polarisation was found to develop below $T_2$, and was later understood to arise from trilinear magneto-structural coupling that creates polar distortions through optimization of \emph{anisotropic} symmetric exchange interactions between the A' and B magnetic sublattices \cite{Feng_2016,Zhao_2020}. Similar effects of magnetically induced electric polarisation have been reported in the related compound BiMn$_3$Cr$_4$O$_{12}$, which shows the same sequence of G-type antiferromagnetic phase transitions at $T_\mathrm{N} = 125~\mathrm{K}$ and $T_2 = 48~\mathrm{K}$. The substitution of Bi for La at the A site \cite{Zhou_2017,Dai_2019,Maia_2023} introduces an additional polar instability originating in the stereo-chemical activity of the Bi$^{3_+}$ cation, which gives rise to an additional proper electric polarisation \cite{Zhou_2017}. The synthesis of PbMn$_3$(Cr$_3$Mn)O$_{12}$ was recently reported \cite{Liu_2021}, in which the heterovalent substitution of Pb$^{2+}$ for La$^{3+}$ at the A site is accommodated by a random 3:1 distribution of Cr$^{3+}$ and Mn$^{4+}$ cations at the B-site, both with the same 3d$^3$, S = 3/2 electronic configuration. Again, PbMn$_3$(Cr$_3$Mn)O$_{12}$ was found to show a series of magnetic phase transitions associated with G-type antiferromagnetic order on both sublattices ($T_\mathrm{N} = 155~\mathrm{K}$ and $T_2 = 81~\mathrm{K}$), however, no ferroelectric polarisation could be found, neither magnetically induced below $T_2$ nor associated with Pb$^{2+}$ stereochemistry. Again, the absolute direction of magnetic moments was not determined.

In this paper, we report the synthesis and characterisation of the quadruple perovskite CaMn$_3$(Cr$_3$Mn)O$_{12}$ (CMCMO). We show that this compound adopts the same G-type magnetic structures reported in the A = La, Bi, and Pb compounds described above (up to the global moment direction which again cannot be determined from powder diffraction). Similar to the case of PbMn$_3$(Cr$_3$Mn)O$_{12}$, we found that CMCMO does not develop an electric polarisation below $T_2$. In light of this result, we propose a phenomenological model of magneto-structural coupling that naturally explains why some such compounds are electrically polar below $T_2$ while others are non-polar and chiral, depending on the global direction of magnetic moments. Importantly, our model implies novel schemes of multiferroic functionality that may be generalised to other material systems.

CaMn$_3$(Cr$_3$Mn)O$_{12}$ was synthesized from a stoichiometric mixture of CaMnO$_3$, Mn$_2$O$_3$, and Cr$_2$O$_3$ by a high-pressure, high-temperature method at about 6 GPa and about 1800 K for 1 hour in Pt capsules. Magnetic susceptibility measurements were performed on a SQUID magnetometer (Quantum Design, MPMS-XL-7T) between 2 and 400 K in a 7 T applied field under field-cooled on warming (FCW) conditions. Specific heat, C$_p$ was recorded on cooling by a pulse relaxation method using a commercial calorimeter (Quantum Design PPMS). Dielectric properties were measured on a Quantum Design PPMS using a \textsc{novocontrol} Alpha-A High Performance Frequency Analyzer between 5 and 300 K on cooling and heating in the frequency range of 100 Hz and 2 MHz. Pyrocurrent measurements were performed on a dense polycrystalline cylindrical pellet of thickness 0.95 mm. Gold contacts of area 17.35 mm$^2$ were applied on either face, and the sample was poled on cooling from 80 to 25 K under a 100 V potential. An oscillatory background signal of frequency $\sim2$ Hz was removed from the data through fast Fourier transform. Neutron powder diffraction data were collected using the WISH diffractometer at ISIS, UK \cite{Chapon2011}. A 640 mg sample was loaded into a cylindrical 3 mm diameter vanadium can, mounted within a closed-cycle refrigerator, and cooled to a base temperature of 10 K. Diffraction data with good counting statistics were collected on warming from 25 to 200 K in 5 K steps, and data with high counting statistics were collected at 10, 100 and 200~K; deep within each magnetic phase. Rietveld refinements were performed using \textsc{fullprof} \cite{Rodriguez-Carvajal1993}, and the \textsc{isotropy} software suite \cite{isotropy,Campbell2006} was used for symmetry analysis.

\begin{figure}
\includegraphics[width=0.49\textwidth]{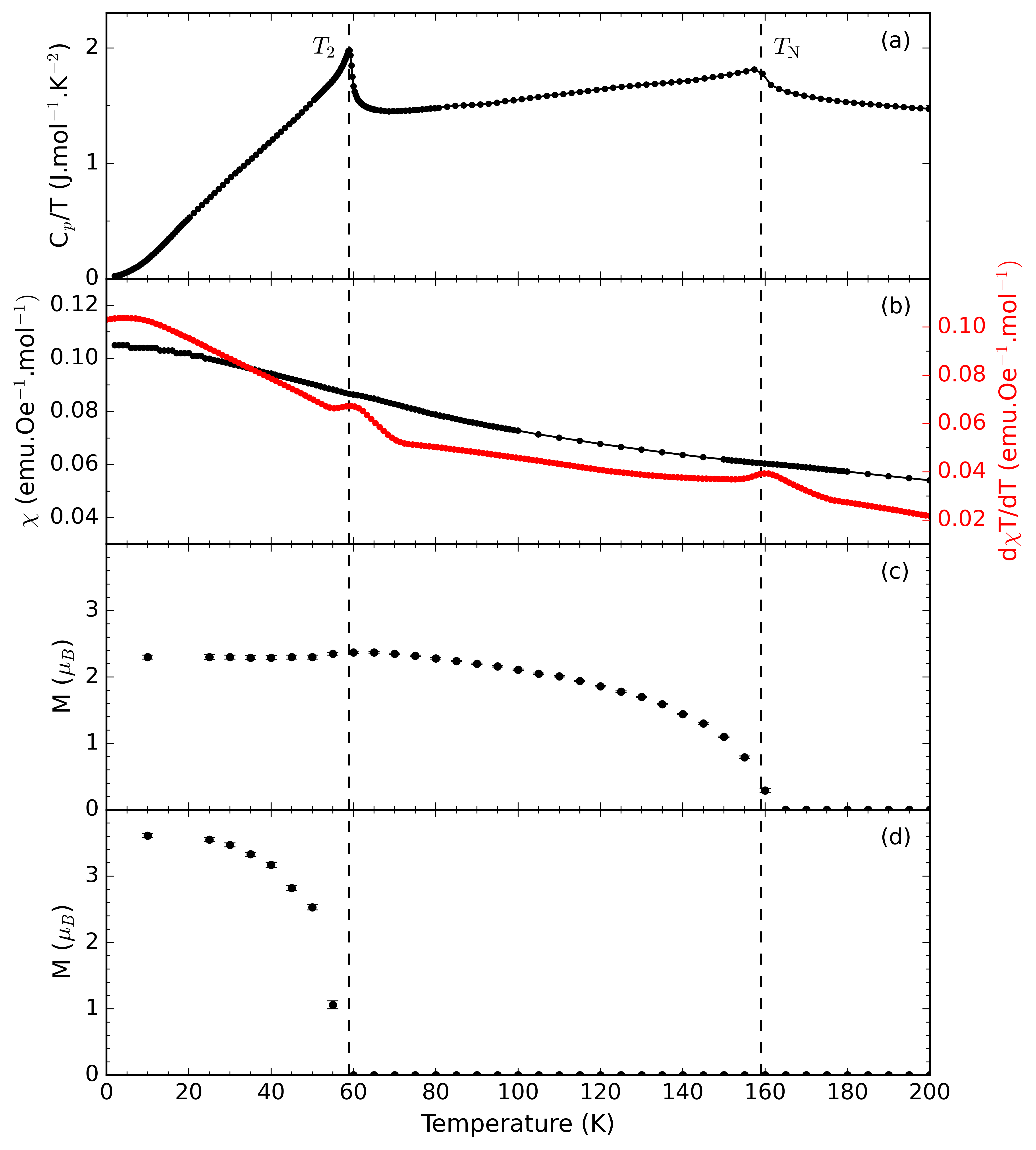}
\caption{a) The specific heat of CaMn$_3$(Cr$_3$Mn)O$_{12}$ measured in zero magnetic field, and b) the magnetic susceptibility, $\chi$, measured in a 7T field on field-cooled-warming (black) and $\mathrm{d}\chi T/\mathrm{d}T$ (red), where $\chi T$ was first interpolated before its gradient determined. c) and d) the average magnetic moment at the B- and A'-sites, respectively. Vertical dashed lines denote the magnetic phase transition.}
\label{FIG::mag_tempdep} 
\end{figure}

The specific heat of CMCMO, measured as a function of temperature, is shown in Figure \ref{FIG::mag_tempdep}a. Two anomalous features indicative of phase transitions were observed at $T_2 = 59~\mathrm{K}$ and $T_\mathrm{N} = 159~\mathrm{K}$. The magnetic susceptibility, $\chi$, is shown in Figure \ref{FIG::mag_tempdep}b. While only slight changes in gradient can be seen at $T_2$ and $T_\mathrm{N}$, a plot of $\mathrm{d}\chi T/\mathrm{d}T$ clearly shows anomalous features at these two temperatures, indicating that both phase transitions involve magnetic degrees of freedom.

\begin{figure}
  \includegraphics[width=0.49\textwidth]{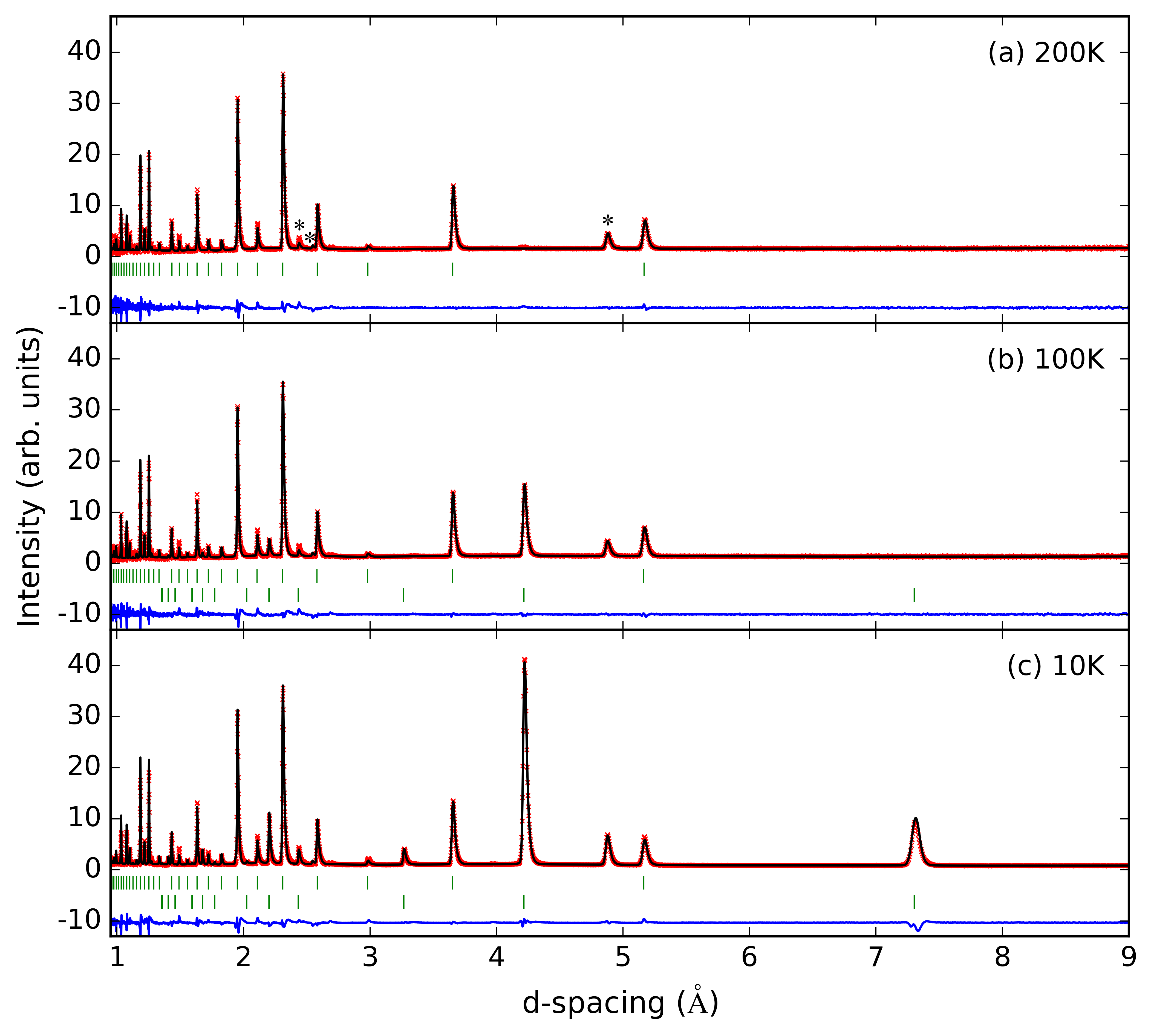}
  \caption{Neutron powder diffraction data (red crosses) collected in banks 2+9 of the WISH diffractometer (average $2\theta = 58.3^\circ$) measured in a) the paramagnetic phase above $T_\mathrm{N}$, b) between $T_\mathrm{N}$ and $T_2$, and c) below $T_2$. Fits to the data are shown as solid black lines, the differences between data and fit are shown as solid blue lines, and the position of nuclear and magnetic Bragg peaks are marked by top and bottom green tick marks, respectively. Significant MnCr$_2$O$_4$ impurity peaks are marked by asterisks in (a).}
  \label{FIG::NPD} 
\end{figure}

Neutron powder diffraction experiments were performed to characterise the magnetic state within each phase. Figure \ref{FIG::NPD}a shows the diffraction data collected above $T_\mathrm{N}$ at 200 K, taken to be the paramagnetic reference. Almost all Bragg peaks could be indexed by the cubic quadruple perovskite $Im\bar3$ lattice, while a small number of weak peaks corresponded to an MnCr$_2$O$_4$ impurity phase. A structural model based on the crystal structure of PbMn$_3$(Cr$_3$Mn)O$_{12}$ was refined against data collected in 8 detector banks of the WISH instrument (2+9, 3+8, 4+7, and 5+6, centred at $2\theta = 58.3^\circ$, $90.0^\circ$, $121.7^\circ$, and $152.8^\circ$, respectively). A reasonable goodness of fit was achieved ($R_\mathrm{Bragg} = 5.1\%$, $R_\mathrm{w} = 9.2\%$), and the structural parameters are given in Table SI of the Supplemental Material \cite{SM}. Refinement of the crystal structure against data collected down to 10 K showed that the sample remained cubic at all measured temperatures. The temperature dependence of the lattice parameter, and A'-O and B-O bond lengths are plotted in Figure S1 \cite{SM}.

\begin{figure}
  \includegraphics[width=0.49\textwidth]{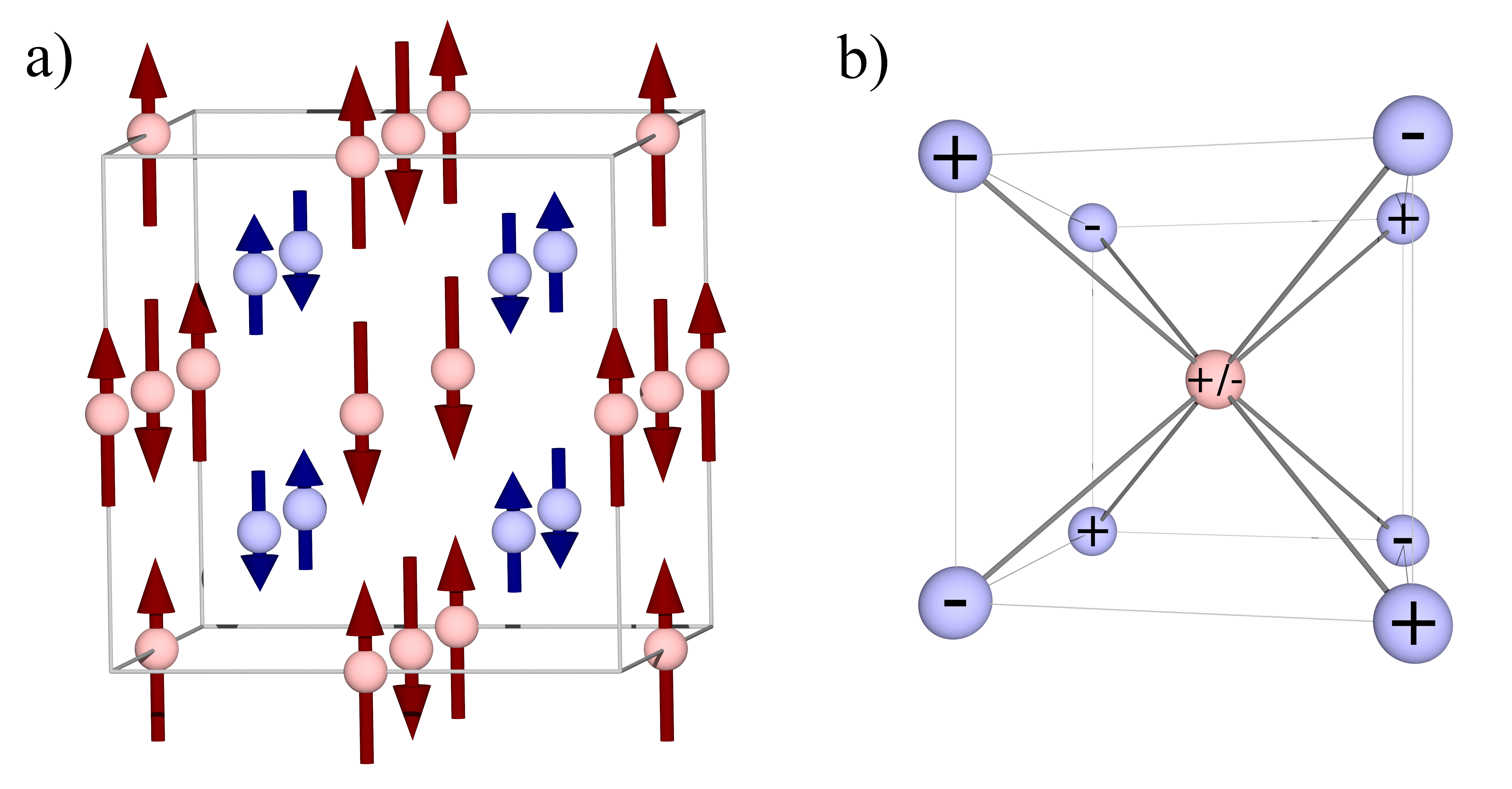}
  \caption{a) Ground state magnetic structure of CaMn$_3$(Cr$_3$Mn)O$_{12}$. Moments are arbitrarily drawn parallel to a unit cell axis. b) Schematic demonstrating exact magnetic frustration of symmetry equivalent A'-B interactions (dark grey lines). A' and B sites drawn in red and blue, respectively.}
  \label{FIG::mag_structure} 
\end{figure}

On cooling below $T_\mathrm{N}$, new magnetic Bragg peaks appeared that could be indexed with propagation vector $\mathbf{k}=(1,1,1)$ (Figure \ref{FIG::NPD}b), which indicates the onset of long-range magnetic order within the same I-centred paramagnetic unit cell, but with $I$-centring vectors connecting anti-aligned moments (anti-translation). More specifically, the magnetic Bragg peaks satisfied the reflection conditions $h+k+l = 2n+1$, $h,k,l$ all odd, which uniquely identifies G-type antiferromagnetic order of the B-site sublattice. On further cooling below $T_\mathrm{2}$ the magnetic Bragg peaks showed an anomalous change in intensity, concomitant with the development of additional magnetic Bragg peaks that also indexed with propagation vector $\mathbf{k}=(1,1,1)$. These changes satisfy the more relaxed reflection conditions, $h+k+l = 2n+1$, $h,k,l$ all odd or $h,k,l$ two even one odd, which are consistent with the onset of G-type antiferromagnetic order of the A' sublattice. This magnetic structure determination was confirmed by Rietveld refinement against data measured at 10 K and 100 K (Figures \ref{FIG::NPD}b and \ref{FIG::NPD}c). While the global moment directions could not be determined due to effects of powder averaging, all ordered moments were found to be collinear in both phases. The temperature dependence of the refined magnetic moment for both B-site and A'-site sublattices is presented in Fig.\ref{FIG::mag_tempdep}(c) and Fig.\ref{FIG::mag_tempdep}(d), respectively, and the magnetic structure is shown schematically in Figure \ref{FIG::mag_structure}a.

The refined magnetic structures are the same as those found in the other A$^{x+}$Mn$_3$(Cr$_{1+x}$Mn$_{3-x}$)O$_{12}$ compounds described in the introduction, up to the global direction of magnetic moments which have not been determined experimentally for any of these compounds. This result suggests that CMCMO may host a magnetically induced electric polarisation below $T_2$, as observed in the A = La and Bi compounds \cite{Wang_2016,Zhou_2017}. The temperature dependence of the dielectric constant of CMCMO is plotted in Figure S2a \cite{SM}. A small anomalous feature can be observed at $T_2$, consistent with magneto-structural coupling at this phase transition. However, a sharp feature consistent with the onset of ferroelectric polarization, as observed in LaMn$_3$Cr$_4$O$_{12}$ \cite{Wang_2016}, is absent. The integrated pyrocurrent signal plotted in Figure S2b \cite{SM} showed evidence for a slowly varying polarization likely due to Maxwell-Wagner type relaxation effects, as is typical for polycrystalline oxides with relatively low resistivity \cite{Terada_16}. However, no anomalous change in polarization could be seen at $T_2$. Taken together, these measurements indicate that CMCMO does not display a magnetically induced electric polarization.

Taking the $Im\bar{3}$ crystal structure as the parent, one can show that the A' and B sublattice G-type magnetic structures transform by different 3D irreducible representations (irreps), mH$_4^+(\eta^c,\eta^a,\eta^b)$ and mH$_4^-(\epsilon^c,\epsilon^a,\epsilon^b)$, respectively. Here, the respective order parameter components are given in parenthesis, where the subscripts denote the direction of magnetic moments of a G-type structure. For example, a B sublattice G-type magnetic structure with moments along $c$ is associated with the 3D order parameter $(\epsilon^c,0,0)$. Given that the A' and B magnetic structures transform by different irreps, they are not coupled in $Im\bar{3}$ symmetry, \emph{i.e.} the B-site magnetic `exchange-field' at any given A-site is exactly equal to zero and the symmetry equivalent A'-B exchange interactions are exactly frustrated, as shown in Figure \ref{FIG::mag_structure}b. This scenario is consistent with the B-sites ordering independently below $T_\mathrm{N}$, due presumably to dominant antiferromagnetic B-B interactions. Below $T_2$ A'-sites order, likely due to antiferromagnetic A'-A' interactions that happen to be lower in energy than the B-B interactions. 

We now identify symmetry breaking atomic displacements that will allow finite exchange coupling between A' and B magnetic sublattices below $T_2$. We require symmetry-adapted displacive modes that form trilinear free energy terms, invariant with respect to the parent $Im\bar{3}$ symmetry, and including both mH$_4^+(\eta^c,\eta^a,\eta^b)$ and mH$_4^-(\epsilon^c,\epsilon^a,\epsilon^b)$ order parameters. Translation and parity invariance immediately imply $\Gamma$-point, non-centrosymmetric displacive modes. The full $\Gamma$-point representation of atomic displacements of the $Im\bar{3}$ structure decomposes into 6 physically real irreducible representations, three of which are antisymmetric with respect to inversion; $\Gamma_1^-(\delta)$, ${\Gamma_2^-\oplus\Gamma_3^-}(\chi_1,\chi_2)$, and $\Gamma_4^-(\xi_1,\xi_2,\xi_3)$, where order parameter components are given in parenthesis. Systematic searches using the \textsc{isotropy} software tool found five trilinear coupling invariants of the required form (which we label I, IIa, IIb, IIIa, or IIIb): 
\begin{equation}
\label{EQN::I}
\mathcal{I}_\mathrm{I}=\delta(\eta^a\epsilon^a + \eta^b\epsilon^b + \eta^c\epsilon^c),
\end{equation}
\begin{equation}
\label{EQN::IIa}
\mathcal{I}_\mathrm{IIa}=\chi_2\big(\eta^c\epsilon^c - \tfrac{1}{2}\left(\eta^a\epsilon^a +\eta^b\epsilon^b \right)\big) + \tfrac{\sqrt{3}}{2}\chi_1(\eta^a\epsilon^a - \eta^b\epsilon^b),
\end{equation}
\begin{equation}
\label{EQN::IIb}
\mathcal{I}_\mathrm{IIb}=\chi_1\big(\eta^c\epsilon^c - \tfrac{1}{2}\left(\eta^a\epsilon^a +\eta^b\epsilon^b \right)\big) - \tfrac{\sqrt{3}}{2}\chi_2(\eta^a\epsilon^a - \eta^b\epsilon^b),
\end{equation}
\begin{equation}
\label{EQN::IIIa}
\mathcal{I}_\mathrm{IIIa}=\left(\xi_1\eta^b\epsilon^c + \xi_2\eta^c\epsilon^a + \xi_3\eta^a\epsilon^b\right),
\end{equation}
and,
\begin{equation}
\label{EQN::IIIb}
\mathcal{I}_\mathrm{IIIb}=\left(\xi_1\eta^c\epsilon^b + \xi_2\eta^a\epsilon^c + \xi_3\eta^b\epsilon^a\right).
\end{equation}
$\mathcal{I}_\mathrm{I}$, $\mathcal{I}_\mathrm{IIa}$, and $\mathcal{I}_\mathrm{IIb}$ are symmetric under exchange of A' and B sublattice magnetic order parameters, \emph{i.e.} $\eta_i \leftrightarrow \epsilon_i$, while $(\mathcal{I}_\mathrm{IIIa}+\mathcal{I}_\mathrm{IIIb})$ and $(\mathcal{I}_\mathrm{IIIa}-\mathcal{I}_\mathrm{IIIb})$ are symmetric and antisymmetric, respectively. One can show that all symmetric terms are optimized for collinear A' and B magnetic structures, as observed experimentally. Conversely, the antisymmetric case, $\mathcal{I}_\mathrm{IIIa}-\mathcal{I}_\mathrm{IIIb}$, is exactly zero for collinear sublattices.

$\mathcal{I}_\mathrm{I}$ is isotropic in the magnetic degrees of freedom, and facilitates coupling between A' and B sublattices by inducing a displacive mode that transforms as $\Gamma_1^-$. In this case the crystal point group symmetry is lowered from $m\bar{3}$ to $23$; a non-polar, chiral group. $\mathcal{I}_\mathrm{IIa}$ and $\mathcal{I}_\mathrm{IIb}$ are anisotropic in the magnetic degrees of freedom, optimized for moments along $\mathbf{a}$, $\mathbf{b}$, or $\mathbf{c}$, and exactly zero for moments parallel to the cubic body diagonals. This invariant facilitates coupling between A' and B sublattices by inducing a displacive mode that transforms as ${\Gamma_2^-\oplus\Gamma_3^-}$, which lowers the crystal point group symmetry to non-polar, chiral $222$. The direction of the ${\Gamma_2^-\oplus\Gamma_3^-}$ order parameter, $(\chi_1,\chi_2)$, depends on the relative magnitudes of the respective free energy coefficients and the direction of the magnetic moments \cite{SM}. The symmetric sum $(\mathcal{I}_\mathrm{IIIa}+\mathcal{I}_\mathrm{IIIb})$ is minimised for moments aligned with the body diagonals, and is zero for moments parallel to $a$, $b$, or $c$; exactly opposite to the behaviour of $\mathcal{I}_\mathrm{IIa}$ and $\mathcal{I}_\mathrm{IIb}$. Here, coupling between A' and B sublattices is introduced by a $\Gamma_4^-$ displacive mode, which results in a polar crystal point group. Moments along [111], \emph{i.e.} $\eta^a=\eta^b=\eta^c$ and $\epsilon^a=\epsilon^b=\epsilon^c$, stabilize a distortion with order parameter $(\xi,\xi,\xi)$ \cite{SM} giving the polar point group 3, which allows a ferroelectric polarisation parallel to the magnetic moments. The above free energy terms can be mapped onto A'-B exchange terms in the spin Hamiltonian, as shown in detail in the Supplemental Material \cite{SM}. The fact that the energy terms II and III are optimised by different order parameter directions (by different direction of magnetic moments in the lattice), and that one type of coupling vanishes when the other is optimal reveals their competing nature. This provides an elegant interpretation of the magneto-structural phenomena observed in the cubic $Im\bar{3}$ perovskites. The free-energy term I selects the collinear alignment of the A' and B sublattices in the ground state, but its isotropic nature cannot choose the global direction of the spins. This direction is chosen by the competing anisotropic terms II and III. While the global direction of magnetic moments has not been determined in any of the following materials, we propose that IIIa and IIIb magneto-structural coupling selects the polar ground state observed in LaMn$_3$Cr$_4$O$_{12}$ and BiMn$_3$Cr$_4$O$_{12}$ with moments parallel to $\langle 111 \rangle$, while the IIa and IIb coupling stabilises the non-polar chiral state realised in CaMn$_3$(Cr$_3$Mn)O$_{12}$ and PbMn$_3$(Cr$_3$Mn)O$_{12}$ with moments parallel to $\langle 100 \rangle$. The predominance of one coupling scheme over another will be determined by microscopic characteristics of these materials, in particular by the elastic energy cost associated with polar or chiral distortions. 

In summary, we have shown that CaMn$_3$(Cr$_3$Mn)O$_{12}$ undergoes two magnetic phase transitions at $T_\mathrm{N} = 159~\mathrm{K}$ and $T_2 = 59~\mathrm{K}$, below which the B and A’ sublattice moments order with collinear G-type antiferromagnetic structures, respectively.  The same sequence of magnetic phase transitions was observed in LaMn$_3$Cr$_4$O$_{12}$, where a magnetically induced electric polarization was also observed below $T_2$. We found that this multiferroic electric polarization is absent in CaMn$_3$(Cr$_3$Mn)O$_{12}$. To account for this difference, we present a phenomenological model of trilinear magneto-structural coupling that describes secondary, symmetry breaking structural distortions necessarily occurring at $T_2$ to facilitate coupling between the two sublattices. Our model predicts that if magnetic moments lie along the body diagonal a polar crystal structure is stabilized with electric polarization parallel to the body diagonal. To the contrary, if magnetic moments are aligned along the primary cubic axes, a chiral but non-polar structure results. This phenomenology is an exemplary manifestation of the general principle that in high symmetry crystals the direction of magnetic moments can determine the global symmetry of the system via magneto-structural coupling.  Importantly, the presence of competing polar and chiral ground states implies an intriguing paradigm of multiferroic behavior. A spontaneous electric polarization can be switched on and off at fixed temperature just by changing the direction of the antiferromagnetic magnetic moments (N$\mathrm{\acute{e}}$el vector). Conversely, and of great significance to antiferromagnetic spintronics, the direction of the N$\mathrm{\acute{e}}$el vector can be controlled with an external electric field. Finally, we stress that our phenomenological model may be generalized to other crystalline systems that support similar network topologies of magnetic frustration for which magnetic interactions between sublattices are prohibited by spatial inversion.

\begin{acknowledgements}

\end{acknowledgements}

\bibliography{ref}

\end{document}